\documentstyle[12pt]{article}
\newcommand{\beq}{\begin{equation}}
\newcommand{\eeq}{\end{equation}}
\newcommand{\beqa}{\begin{eqnarray}}
\newcommand{\eeqa}{\end{eqnarray}}
\newcommand{\beqan}{\begin{eqnarray*}}
\newcommand{\eeqan}{\end{eqnarray*}}
\newcommand{\half}{{{1}\over{2}}}
\newcommand{\ihalf}{{{i}\over{2}}}
\newcommand{\quar}{{{1}\over{4}}}
\newcommand{\la}{\lambda}

\newcommand{\bra}{\langle 0|}
\newcommand{\ket}{|0\rangle}
\newcommand{\id}{{1}\hspace{-0.3em}\rm{I}}

\newcommand{\si}{\sigma}
\newcommand{\ep}{\epsilon}
\newcommand{\cer}{\tilde{C}}
\newcommand{\br}{\tilde{B}}

\newcommand{\ar}{\tilde{A}}
\newcommand{\dr}{\tilde{D}}

\newcommand{\sump}{\sum_{I,II,III}\,\hspace{-0.9em}^\prime}
\begin{document}
\baselineskip=18pt
\pagenumbering{arabic}
\parskip1.5em
\thispagestyle{empty}
\begin{flushright}BN--TH--97--09\\
\end{flushright}
\vskip4em
\begin{center}
{\Large{\bf Multi--Magnon Scattering in the Ferromagnetic \\
\bigskip
XXX--Model with Inhomogeneities}}\vskip2em
by\vskip2em
T.-D. Albert\hspace{3em}R. Flume\hspace{3em}K. Ruhlig\\\vskip3em
{\sl Physikalisches Institut der Universit\"at Bonn}\\
{\sl Nu\3allee 12}\\
{\sl D--53115 Bonn}\\
{\sl Germany}
\end{center}
\vskip2em
\begin{abstract}
We determine the transition amplitude for multi--magnon scattering induced through an inhomogeneous distribution of the coupling constant in the ferromagnetic XXX--model. The two and three particle amplitudes are explicitely calculated at small momenta. This suggests a rather plausible conjecture also for a formula of the general n--particle amplitude.
\end{abstract}\vskip2em
e-mail: t-albert@avzw01.physik.uni-bonn.de\\
\hspace*{1.3cm}flume:\hspace{2em}unp06d@IBM.rhrz.uni-bonn.de \\
\hspace*{1.3cm}ruhlig@avzw01.physik.uni-bonn.de
\vfill\eject\setcounter{page}{1}
\section{Introduction}
We want to report in this article the calculation of transition amplitudes of multi-magnon scattering in the ferromagnetic Heisenberg XXX--chain with an inhomogeneous distribution of the coupling constant.\\
The Hamiltonian of the model under consideration is given by
\beqan
\hspace*{4.4cm}H&=&H_{hom}+H_{inh}\hspace{8.2cm}(1)\\
H_{hom}&=&{J\over{4}}\sum^N_{n=-N}[\vec{\si}_n\vec{\si}_{n+1}-\id]\label{ham}\hspace{6.55cm}(1\,a)\\
H_{inh}&=&\quar\sum^N_{n=-N}z_n[\vec{\si}_n\vec{\si}_{n+1}-\id]\label{haminh}\hspace{6.1cm}(1\,b)\\
&&\vec{\si}\cdot\vec{\si}=\sum_{a=1}^3 \si^a\cdot\si^a
\eeqan
\setcounter{equation}{1}
where $\si^a_i$ denotes the Pauli matrices operating in quantum spaces $V_i;\,\,i=-N,\dots,N$ attached to a one-dimensional lattice with $2N$ sites.\\
We choose in (\ref{ham}) the ferromagnetic sign of the coupling constant $(J<0)$ and assume the inhomogeneous piece $H_{inh}$ to be a small perturbation of the homogeneous part $H_{hom}$, that is, we stipulate for the locally varying couplings $z_i$
\beqan
|z_i|\ll |J|\,.
\eeqan
The homogeneous XXX-chain is, as a prototype of an integrable model, one of the most thoroughly studied one-dimensional spin models.\\
A mathematically rigorous analysis of the model has been provided by Babbitt and Thomas \cite{bab}. For a treatment of the XXX--model in the framework of the algebraic Bethe Ansatz (ABA) cf. references \cite{takh}--\cite{bog}. It is found that the complete spectrum of the model is formed by quasi-particles, here called magnons, and bound states of magnons, the so-called string states. The integrability of the model implies that the interaction between magnons and strings is of a particularly simple structure. It is characterized by the following features \cite{zam}:
\begin{itemize}\vspace{-0.5cm}
\item multi-particle scattering factorizes into two-particle amplitudes
\item the string states are absolutely stable bound states.
\end{itemize}
It follows from these properties that neither genuine multi-particle scattering takes place (with a non-trivial reshuf\/f\/ling of the particle momenta) nor does a break-up of the bound states occur.\par
We will make use of the technique of the ABA \cite{takh}--\cite{bog}.
A first step in this direction is to embed the Heisenberg spin model (1a) into a family of vertex models. The latter models are defined through a monodromy matrix $T(\la)$ depending on a spectral parameter $\la$
\beq
T(\la)=L_{N}(\la)\cdots L_{-N}(\la)\label{monod}
\eeq
with the local ``Lax-operators'' $L_n$ given by
\beq
L_n(\la)=\half \left [2 i \la \id_0\otimes \id_n +\vec{\si}_{0}\otimes\vec{\si}_{n} \right ]\label{lax}
\eeq
The unit operator $\id_0$ and the Pauli matrices $\sigma_0$ act in an auxiliary two-dimensional space, while $\id_n$ and  $\sigma_n$ act in the quantum space $V_n$. The spin chain model (1a) emerges as the logarithmic derivative of the vertex model monodromy matrix 
\beqan
H=-J\ihalf{{d\,\ln(tr_0\, T(\la))}\over{d\,\la}}\Big|_{\la=0}-J{2N\over 2}\id= {J\over 4}\sum^N_{n=-N}[\vec{\si}_n\vec{\si}_{n+1}-\id]
\label{hamil}
\eeqan
with $tr_0$ denoting the trace with respect to the auxiliary space.\\
The integrability of the vertex models and therewith also the integrability of the XXX-model is based on the fact that there is a c-number matrix $R=R(\la-\mu)$, s.t. the Yang-Baxter-Faddeev-Zamolodchikov (YBFZ) relation
\beq
R(\la-\mu)\,L_i(\la)\otimes L_i(\mu)=L_i(\mu)\otimes L_i(\la)\,R(\la-\mu)\label{lcr}
\eeq
is satisfied. $R=R(\la-\mu)$ is in the case at hand given by 
\beq
R(\la-\mu)=\left (\begin{array}{cccc} f(\mu{,}\la) & 0 &0 & 0\\ 0 & g(\mu{,}\la) & 1 & 0 \\ 0 & 1 & g(\mu{,}\la) & 0 \\0 & 0      & 0      & f(\mu{,}\la) \\
\end{array}\right )\label{rmatrix}
\eeq
with
\beqan
f(\mu{,}\la)=1+{{i c}\over{\mu-\la}}\mbox{ , and}\qquad g(\mu{,}\la)={{i c}\over{\mu-\la}}.
\eeqan
The parameter $c$ is set to unity in the following, meaning that the spectral parameter is taken as a dimensionless entity.
The local relation (\ref{lcr}) induces the global relation
\beq
R(\la-\mu)\left (T(\la)\otimes T(\mu)\right )=\left (T(\mu)\otimes T(\la)\right )R(\la-\mu)
\label{gcr}
\eeq
which might be considered here as hallmark of integrability.\\
The YBFZ-relations can be maintained in certain inhomogeneous generalizations of the above models. One possibility is to choose different representations for different sites of the lattice. Such cases have been analyzed in references \cite{deveg2}--\cite{doerfel}. A conceptually simpler possibility, noticed in \cite{devega, bog}, consists in attaching local parameters $z_i$ to the local Lax-operator, that is one substitutes $L_i(\la)$ by $L_i(\la-z_i)$ and obtains then also a modified monodromy matrix
\beq
T(\la,\{z_i\})=T(\la;z_{-N}\ldots,z_N)=\prod_{i=N}^{-N} L_i(\la-z_i)\label{monodinh}\,\,.
\eeq
One may easily see that the YBFZ relations remain intact,
\beq
R(\la-\mu)\left (T(\la,\{z_i\})\otimes T(\mu,\{z_i\})\right )=\left (T(\mu,\{z_i\})\otimes T(\la,\{z_i\})\right )R(\la-\mu).
\label{gcrinh}
\eeq
It should be noted that equation (\ref{gcrinh}) only holds in general if one specifies for $T(\la,\{z_i\})$ and $T(\mu,\{z_i\})$ the same distribution of local parameters $\{z_{-N},\ldots,z_N\}$. The physics of the model on the other hand appears to be invariant under permutations of the parameters. This is a consequence of the fact that the BA equations, to be mentioned shortly in the following section, which provide the spectrum of the eigenstates, are insensitive to these permutations. It is true (modulo some inessential caveats) that the order of the different representations along the lattice, as mentioned above, is for the same reason irrelevant. We therefore believe that genuine effects of inhomogenities can only be realized outside the class of integrable models.\\It is an easy undertaking to arrive from the inhomogeneous vertex model (\ref{monodinh}) at the inhomogeneous spin chain (1b). Let us make for this purpose the specifications 
\beq
\la\rightarrow\epsilon\,\la\hspace{4em}z_j\rightarrow\epsilon z_j\,\,.
\eeq
The inhomogeneous Heisenberg magnet (1b) is recovered as the logarithmic derivative of the vertex model
\begin{eqnarray}
H&=&-J{{i}\over2}{d\over{d\ep}}ln(tr_0\,T(\ep\la,\{\ep z_i\}))\bigg{|}_{\ep=0}-{J\over 2}\sum^N_{n=1}(\la-z_n)\id \nonumber \\
&& \nonumber \\
&=& {J\over 4}\sum^N_{n=1}(\la-z_n)[\vec{\si}_n\vec{\si}_{n+1}-\id]=:\la H_0-H_1
\label{hamstoer}.
\end{eqnarray}
By taking the derivative with respect to a parameter $\epsilon$ which parametrizes different distributions, one expects to leave the realm of integrability. We will confirm this expectation by evaluating non-vanishing irreducible multi-particle scattering amplitudes which are supposed to vanish identically in integrable models.\\
We will restrict our considerations to the scattering of elementary magnons (the incorporation of string states is technically definitely much more cumbersome). A simplifying aspect of the problem can be found in that the inhomogeneous perturbation respects the same global $SU(2)$ invariance as the homogeneous term. This implies magnon number conservation.\par
The plan of the paper is as follows: in the subsequent section we recall some ingredients of the ABA, discuss the thermodynamic limit and introduce the so-called multisite formalism, taken from \cite{kor}. In section III we evaluate in first order perturbation theory multi-particle amplitudes at small momenta. Of crucial importance to achieve this goal will be the representation of form factors as deduced in \cite{kor}.
The technical tools to be applied in our analysis are approximately the same as those used in \cite{sal} for perturbative calculations in antiferromagnetic environments (while of course our calculations are simpler and more simple--minded). The concluding section is devoted to a qualitative discussion of other kinematical regions of multi-particle scattering and a summary. In an appendix we report a perturbative calculation of the spectrum of low lying states which is conceptually not related to the theme of the bulk of the paper but on a technical level rather similiar to the calculations in section III.
\section{Basics of Bethe Ansatz}
\subsection{Algebraic Bethe Ansatz}
We collect here for the sake of self-consistency of the paper some of the basic aspects of the ABA \footnote{For a more thorough introduction see \cite{takh, faddeev}.}.\\
Let the monodromy matrix (\ref{monod}) be parameterized as
\beq
T(\la)=\left ( \begin{array}{cc} A(\la) & B(\la) \\ C(\la) & D(\la) \end{array} \right ).
\eeq
One deduces from the YBFZ relations (\ref{gcr}) the commutators of the operators $A(\la),\dots,D(\la)$. Of these 16 relations we only list the following 
\begin{eqnarray}
\Big[B(\la),\,B(\mu)\Big] &=& 0\,=\,\Big[C(\la),\,C(\mu)\Big]\nonumber\\
\Big[B(\la),\,C(\mu)\Big] &=&g(\la,\mu)\Big(D(\la)A(\mu)-D(\mu)A(\la)\Big)\nonumber\\
A(\mu)B(\la) &=& f(\mu,\la)B(\la)A(\mu)+g(\la,\mu)B(\mu)A(\la)\nonumber\\
D(\mu)B(\la) &=& f(\la,\mu)B(\la)D(\mu)+g(\mu,\la)B(\mu)D(\la)\nonumber\\
C(\la)D(\mu) &=& f(\la,\mu)D(\mu)C(\la)+g(\mu,\la)D(\la)C(\mu)\nonumber\\
C(\la)A(\mu) &=& f(\mu,\la)A(\mu)C(\la)+g(\la,\mu)A(\la)C(\mu)\,\, .\nonumber\\\label{comm}
\end{eqnarray}
Let $\ket$ denote the state of highest weight with respect to the tensorproduct of $SU(2)$ representations in the configuration space $V=\prod_{\otimes}V_i$. This  state  is annihilated by the operators $C(\la)$ and is an eigenstate of the trace of the transfermatrix $\tau(\la)$
\beqan
\tau(\la)=tr_0 T(\la)&=&A(\la)+D(\la)\nonumber\\
C(\la)\ket &=&0\nonumber\\
\left(A(\la)+D(\la)\right)\ket &=&\left[\left ({i\la +\half}\right)^{2N}+\left ({i\la -\half}\right)^{2N}\right]\ket\,\,.
\eeqan
The ABA renders a representation of the eigenstates of the transfermatrix in terms of the operators $B(\la)$ - being the hermitian conjugates of the operators $C(\la)$ - which act as creation operators of quasi-particles (magnons) on the highest weight state $\ket$.\\
Introducing the notation
\beq
|\Phi(\la_1{,...,}\la_l)\rangle=\prod^l_{n=1}B(\la_n)|0\rangle \label{bwf}
\eeq
one arrives - exploiting the commutation relations (\ref{comm}) - at 
\beq
\left(A(\la)+D(\la)\right)\prod^l_{n=1}B(\la_n)|0\rangle=\Lambda(\la{;}\la_1{,...,}\la_l)\prod^l_{n=1}B(\la_n)|0\rangle+\sum_{k=1}^l\tilde{\Lambda}_k(\la_1{,...,}\la_l)B(\la)\prod_{j\neq k}B(\la_j)\ket
\eeq
with
\begin{eqnarray}
\Lambda(\la{;}\la_1{,...,}\la_l)=\left ({i\la +\half}\right)^{2N}\prod^l_{j=1}{{\la_j-\la-i}\over{\la_j-\la}}+\left ({i \la - \half}\right )^{2N}\prod^l_{j=1}{{\la_j-\la+i}\over{\la_j-\la}},
\end{eqnarray}
and
\begin{eqnarray}
\tilde{\Lambda}_k(\la_1{,...,}\la_l)=g(\la_k,\la)\left[\left ({i\la_k +\half }\right)^{2N}\prod^l_{{j=1}\atop{j\neq k}}{{\la_j-\la_k-i}\over{\la_j-\la_k}}-\left ({i \la_k -\half} \right )^{2N}\prod^l_{{j=1}\atop{j\neq k}}{{\la_j-\la_k+i}\over{\la_j-\la_k}}\right].
\label{BAGLinhom}
\end{eqnarray}
The second bunch of terms on the right handside comes from the exchange term in the commutation relations. The spectral parameters have to be specified s.t. these terms vanish. This gives rise to the Bethe Ansatz equations
\beq
\left({{i\la_k +\half }\over{i\la_k -\half }}\right)^{2N}=\prod^l_{{j=1}\atop{j\neq k}}{{\la_j-\la_k+i}\over{\la_j-\la_k-i}}\,\,.
\eeq
$|\Phi(\la_1{,...,}\la_l)\rangle$ is an eigenstate of the transfermatrix if the last equations are satisfied.\\
The eigenstates can be classified with respect to the eigenvalue of $S^3$:
\beq
S^3|\Phi(\la_1{,...,}\la_l)\rangle=({{2N}\over 2}-l)|\Phi(\la_1{,...,}\la_l)\rangle \,\, .
\eeq
The dual wave functions are given by
\beq
\langle\Phi(\{\la_j\})|=\bra\prod_{j=1}^l B^{\dagger}(\la_j)=(-1)^{l}\bra\prod_{j=1}^N C(\la_j)\,\, .
\eeq
\subsection{Thermodynamical limit}
We introduce suitably normalized operators in order to deal with finite norms of states and finite eigenvalues in the thermodynamical (TD) limit \cite{skly}\footnote{Our prescription differs from the one given in \cite{skly} in an inessential way.}:
\beqa
\ar (\la)=a^{-1}(\la) A(\la)\,&{,}&\,\dr (\la)=d^{-1}(\la) D(\la)\nonumber\\
\br (\la)=a^{-\half}(\la) d^{-\half}(\la) B(\la)\,&{,}&\,\cer (\la)=a^{-\half}(\la) d^{-\half}(\la) C(\la)\label{ro}
\eeqa
with $a(\la)=\alpha(\la)^{2N}=(i\la+\half)^{2N}$ the eigenvalue of the operator $A(\la)$ and $d(\la)=\delta(\la)^{2N}=(i\la-\half)^{2N}$ the eigenvalue of $D(\la)$ with respect to $\ket$. Thus the operators $\ar(\la)$ and $\dr(\la)$ have correspondingly the eigenvalue 1.
Exploiting the relations \footnote{Terms of the form $1\over{x}$ have to be evaluated according to the principle--value prescription.}
\beq
{\lim _{N\rightarrow\infty}} g(\la-\mu)\exp{i(p(\la)-p(\mu))N\over{2}}=-\pi\delta(\la-\mu)\label{delta} 
\eeq
and 
\beqan
{1\over{x}}={1\over{x\pm i\epsilon}}\pm i\pi\,\delta(x)
\eeqan
which hold in the sense of generalized functions  \cite{gelfand} - not pointwise - we obtain in the TD limit the simplified relations
\beqa
\tilde{A}(\la)\tilde{B}(\mu)=f_{-}(\la,\mu)\tilde{B}(\mu)\tilde{A}(\la)\nonumber\\
\tilde{C}(\la)\tilde{D}(\mu)=f_{-}(\la,\mu)\tilde{D}(\mu)\tilde{C}(\la)\label{rcomm}
\eeqa
with $f_{-}(\la,\mu)=1+{i\over{\la-\mu- i\epsilon}}$.\\
One notes, comparing with equation (\ref{comm}), that the exchange terms have dropped out. The BAE can therefore be disregarded in the TD limit (as long as one restricts the attention to the sector of elementary magnons). Normalized asymptotic scattering states are generated by acting with creation operators  $Z(\la)=\tilde{B}(\la)\tilde{A}^{-1}(\la)$ on the vacuum (the highest weight state) and are annihilated by operators $Z^{\dagger}(\la)=-\tilde{D}^{-1}(\la)\tilde{C}(\la)$ \cite{zam}. The action of the operators $\tilde{A}^{-1}$ and $\tilde{D}^{-1}$ are easily deduced from the relations (\ref{rcomm}) and from the fact that the vacuum is an eigenstate with unit eigenvalue of $\tilde{A}(\la)$ and $\tilde{D}(\la)$ and therefore also of $\tilde{A}^{-1}$ and $\tilde{D}^{-1}$.\\
An incoming scattering state is given by 
\beq
Z(\la_1)\ldots Z(\la_n)\ket\label{zwf}
\eeq
if the rapidities are ordered in such a way that $\la_1<\ldots <\la_n$, and represents an outgoing state for $\la_1>\ldots >\la_n$.\\
To relate the incoming to the outgoing states, use has to be made of the relation
\beq
Z(\la)\,Z(\mu)=S(\la,\mu)Z(\mu)\,Z(\la)
\eeq
with $S(\la,\mu)={{f(\la,\mu)}\over{f(\mu,\la)}}$ the two-body S-matrix.
It is easily seen that the n-magnon S-matrix is given as a product of 2-magnon S-matrices.\\
The wavefunctions (\ref{zwf}) are normalized to delta functions with a unit prefactor.\\
\subsection{Multisite formalism}
To evaluate scattering amplitudes in the Born approximation, we have to determine  formfactors \cite{sal} of the type
\beq
\bra\prod_i^l Z^{\dagger}(\la_i){\cal O}\prod_j^k Z(\la_j)\ket 
\eeq
where the operator ${\cal O}$ is given by ${\cal O}=\sum_{n=-N}^N z_n\vec{\si}_n\vec{\si}_{n+1}\equiv \sum_n{\cal O}_{n,n+1}$. So we are lead to consider matrixelements of the form
\beq
\bra\prod_i^l Z^{\dagger}(\la_i^C){\cal O}_{n,n+1}\prod_j^k Z(\la_j^B)\ket=\prod_{i>j}f^{-1}(\la_j^C,\la_i^C)\,f^{-1}(\la_i^B,\la_j^B)\bra\prod_i^l \tilde{C}(\la_i^C){\cal O}_{n,n+1}\prod_j^k \tilde{B}(\la_j^B)\ket\label{matrix}
\eeq
where the latter identity is a straightforward consequence of the definition of $Z$ and the commutation relation (\ref{rcomm}).\\
The basic strategy for the determination of the r.h.s. of (\ref{matrix}) will consist in decomposing the monodromy matrix into parts as follows:
\beq
T(\lambda)=T(3|\lambda)T(2|\lambda)T(1|\lambda)\label{decomp}
\eeq
\begin{eqnarray*}
T(1|\lambda)&=&L_{n-1}(\lambda)\cdots L_{-N}(\lambda)\nonumber\\
T(2|\lambda)&=&L_{n+1}(\lambda)L_i(\lambda)\nonumber\\
T(3|\lambda)&=&L_N(\lambda)\cdots L_{n+2}(\lambda).
\end{eqnarray*}
The sub--monodromy matrices may be parametrized as $T(\la)$ above
\beqan
T(j|\lambda)=\left( \begin{array}{cc} {A}_j(\la) & {B}_j(\la) \\ {C}_j(\la) & {D}_j(\la) \end{array}\right)
\eeqan
with($j$=1,2,3). The product in (\ref{decomp}) is meant to be ordinary matrix multiplication of $2\times 2$ matrices. The $T(j|\la)$ fullfill the global YBFZ commutation relation seperately, acting on the vector space with highest weight $\ket_j$. The highest weight state of the total space is given as a tensorproduct
\beq
|0\rangle=|0\rangle _3\otimes\ket _2\otimes\ket _1\,\, .
\eeq
Using the commutation relations, the operators $A_j$, $D_j$, which appear if the $B(\la)$ are expressed through operators of the subspaces, can be commuted through to the vacuum. This yields the so--called multisite formula \cite{bog}
\begin{eqnarray}  
\prod^{l^B}_{j=1}&&\hspace{-1.5em} B(\lambda^B_j)\ket=\sum_{\{\lambda^{BI}\}\cup\{\lambda^{BII}\}\cup\{\lambda^{BIII}\}=\{\lambda^B\}}\,\prod^{l^B_1}_{j_B\in I}\,\prod^{l^B_2}_{k_B\in II}\,\prod^{l^B_3}_{l_B\in III}\,K_B \nonumber \\
&& \nonumber\\
&& \hspace{-1em}B_3(\lambda^{BIII}_{l_B})\ket_3\otimes B_2(\lambda^{BII}_{k_B})\ket_2\otimes B_1(\lambda^{BI}_{j_B})\ket_1\label{msf}
\end{eqnarray}
with\beqan
K_B&&\hspace{-1.5em}=a_2(\lambda^{BI}_{j_B})d_1(\lambda^{BII}_{k_B})a_3(\lambda^{BI}_{j_B})d_1(\lambda^{BIII}_{l_B})a_3(\lambda^{BII}_{k_B})d_2(\lambda^{BIII}_{l_B})\nonumber\\
&& \nonumber \\
&&\hspace{-1.5em} \times f(\lambda^{BI}_{j_B}{,  }\lambda^{BII}_{k_B})f(\lambda^{BI}_{j_B}{,  }\lambda^{BIII}_{l_B}) f(\lambda^{BII}_{k_B}{,  }\lambda^{BIII}_{l_B})\,\, .
\label{Kb}
\eeqan
The summation in (\ref{msf}) is with respect to the partition of the set of all Bethe parameters $\{\la_j\}$ in 3 disjunct subsets
 $\{\lambda^{BI}\}$, $\{\lambda^{BII}\}$ and $\{\lambda^{BIII}\}$ with
\beqan
card\{\la^{BI}\}=l^B_1{,} \quad card\{\la^{BII}\}=l^B_2{,}\quad card\{\la^{BIII}\}=l^B_3\,{.}
\eeqan
A similar representation can be derived for the dual vector $\bra\prod_{j=1}^l C(\la_j^C)$ 
\begin{eqnarray}  
\bra\prod^{l^C}_{j=1}&&\hspace{-1.5em} C(\lambda^C_j)=\sum_{\{\lambda^{CI}\}\cup\{\lambda^{CII}\}\cup\{\lambda^{CIII}\}=\{\lambda^C\}}\,\prod^{l^C_1}_{j_C\in I}\,\prod^{l^C_2}_{k_C\in II}\,\prod^{l^C_3}_{l_C\in III}\,K_C \nonumber \\
&& \nonumber\\
&& \hspace{-1em}\bra_3 C_3(\lambda^{CIII}_{l_C})\otimes \bra_2 C_2(\lambda^{CII}_{k_C})\otimes \bra_1 C_1(\lambda^{CI}_{j_C})\label{msf2}
\end{eqnarray}
with\beqan
K_C&&\hspace{-1.5em}=d_2(\lambda^{CI}_{j_C})a_1(\lambda^{CII}_{k_C})d_3(\lambda^{CI}_{j_C})a_1(\lambda^{CIII}_{l_C})d_3(\lambda^{CII}_{k_C})a_2(\lambda^{CIII}_{l_C})\nonumber\\
&& \nonumber \\
&&\hspace{-1.5em} \times f(\lambda^{CII}_{j_C}{,  }\lambda^{CI}_{k_C})f(\lambda^{CIII}_{j_C}{,  }\lambda^{CI}_{l_C}) f(\lambda^{CIII}_{k_C}{,  }\lambda^{CII}_{l_C})\,\, .
\label{Kc}
\eeqan
Inserting (\ref{msf}) and (\ref{msf2}) into (\ref{matrix}) we obtain 
 \begin{eqnarray}
&&\hspace{-1.5em}\langle 0|\prod^{l}_{j=1}C(\lambda^C_j)\,{\cal O}_{n,n+1}\,\prod^{l}_{k=1}B(\lambda^B_k)|0\rangle= \sum_{I,II,III}\prod_{I,II,III}\,\prod_{I\leq J<K\leq III}\,a_J(\la^C_K)d_K(\la^C_J)a_K(\la^B_J)d_J(\la^B_K)\nonumber \\
&&\nonumber\\
&&\hspace{-1.5em}\times \quad f(\la^C_J,\la^C_K)\,f(\la^B_K,\la^B_J)\,{\cal S}^{(I)}_{l_1}(\{\la^C_{I}\},\{\la^B_{I}\})\,{\cal S}^{(III)}_{l_3}(\{\la^C_{III}\},\{\la^B_{III}\})\,\,\bra C_2(\la^C_{II}){\cal O}_{n,n+1}\,B(\la^B_{II})\ket\nonumber\\
\label{mael}
\end{eqnarray}
with
${\cal S}_{l_i}^j(\{\la_j^C\},\{\la_k^B\})=_j\bra\prod_i C(\la^C_i)\prod_i B(\la^B_i)\ket_j$ being the scalar product in the j-th space. The cardinality of the partition sets $\{\la^C_i\}$ and $\{\la^B_i\}$ are equal to each other. Matrixelements with $card\{\la_j^C\}\neq card\{\la_k^B\}$ vanish.\\
Taking into account the normalization of the operators $\tilde{B}$ and $\tilde{C}$ relative to $B$ and $C$ we arrive at the following expression
\begin{eqnarray}
&&\hspace{-1.5em}\langle 0|\prod^{l}_{j=1}\cer(\lambda^C_j)\,{\cal O}_{n,n+1}\,\prod^{l}_{k=1}\br(\lambda^B_k)|0\rangle=\nonumber\\
&&\sum_{{I},{II},{III}}\prod_{I,II,III}\bra C_2(\la^C_{II}){\cal O}_{n,n+1}B_2(\la^B_{II})\ket\prod_{I\leq J<K\leq III}\,f(\la^C_J,\la^C_K)\,f(\la^B_K,\la^B_J)\nonumber \\
&& \nonumber \\
&&\hspace{-1.5em}\times \quad \alpha\delta(\la^C_I)^{-{N_1\over{2}}} \alpha\delta(\la^C_{III})^{-{N_3\over{2}}} \alpha\delta(\la^B_I)^{-{N_1\over{2}}} \alpha\delta(\la^B_{III})^{-{N_3\over{2}}} \alpha\delta(\la^B_{II})^{-1} \alpha\delta(\la^C_{II})^{-1}\nonumber \\
&& \nonumber \\
&&\hspace{-1.5em}\times \quad \left[{r(\la^B_I)\over{r(\la^C_I)}}\right]^{{N_3\over{2}}+1}\left[{r(\la^C_{III})\over{r(\la^B_{III})}}\right]^{{N_1\over{2}}+1}\left[{r(\la^B_{II})\over{r(\la^C_{II})}}\right]^{{N_3-N_1}\over{2}}
{\cal S}^{(I)}_{l_1}(\{\la^C_I\},\{\la^B_I\})\,{\cal S}^{(III)}_{l_3}(\{\la^C_{III}\},\{\la^B_{III}\})\label{renme}\nonumber\\
\end{eqnarray}
where $\alpha\delta(\la)\equiv\alpha(\la)\delta(\la)$ and $r(\la)\equiv{\alpha(\la)\over{\delta(\la)}}$.\\
One has the following recursion relation for the scalarproducts \cite{kor}:
\beqan
{\cal S}_{l}(\{\la^{C}_j\},\{\la^{B}_k\})=&&a(\la_1^C)\sum_{n=1}^{l}d(\la_n^B)g(\la_1^C,\la_n^B)\prod_{j\neq 1}^{l}g(\la_1^C,\la_j^C)\prod_{k\neq n}^{l}g(\la_k^B,\la_n^B){\cal S}_{l-1}(\hat{a}_1(\la),\hat{d}_1(\la))+\nonumber\\
&& \nonumber \\
&&d(\la_1^C)\sum_{n=1}^{l}a(\la_n^B)g(\la_n^B,\la_1^C)\prod_{j\neq 1}^{l}g(\la_j^C,\la_1^C\la_1^C)\prod_{k\neq n}^{l}g(\la_n^B,\la_k^B){\cal S}_{l-1}(\hat{a}_2(\la),\hat{d}_2(\la))\label{recrel}\nonumber\\
\eeqan
with $\hat{a}_1(\la)=a(\la)h(\la,\la^B_n){,}\quad\hat{a}_2(\la)=a(\la)h(\la,\la^C_1)$ and $\hat{d}_1(\la)=d(\la)h(\la^C_1,\la){,}$\\
$\hat{d}_2(\la)=d(\la)h(\la^B_n,\la)$, while $h(\la,\mu)={{g(\la,\mu)}\over{f(\la,\mu)}}=1+{{\la-\mu}\over{i}}$. We have quoted here on the r.h.s. the functional dependence of the scalar products on the vacuum eigenvalues which have changed going from the l.h.s. to the r.h.s. from $a(\la)$ to $\hat{a}(\la)$ and $d(\la)$ to $\hat{d}(\la)$ respectively, which makes the solution of the recursion relation difficult in general. The two-term recursion relation simplifies in the TD limit, if we concentrate on the irreducible part of the amplitude.\\
We get for the normalized scalarproduct in the limit $N\rightarrow\infty$:
\beqa
&&{\lim _{N\rightarrow\infty}}\prod_{CI}^{l_1}\prod_{BI}^{l_1}\left[{r(\la^B_I)\over{r(\la^C_I)}}\right]^{{N_3\over{2}}+1}\alpha\delta(\la^C_I)^{-{N_1\over{2}}} \alpha\delta(\la^B_I)^{-{N_1\over{2}}}{\cal S}^{(I)}_{l_1}(\{\la^C_I\},\{\la^B_I\})= \nonumber\\
&&\hspace{2.5em}\left\{\sum_{n_1=1}^{l_1}\left[{r(\la_{n_1}^{BI})\over{r(\la_{n_1}^{CI})}}\right]^{{{N_3-N_1}\over{2}}+1}g(\la_1^{CI},\la_{n_1}^{BI})\prod_{{j_1}\neq 1}^{l_1}g(\la_1^{CI},\la_{j_1}^{CI})\prod_{k_1\neq n_1}^{l_1}g(\la_{k_1}^{BI},\la_{n_1}^{BI})\right.\nonumber\\
&&\hspace{5.5em}\left.-\pi\sum_{n_1=1}^{l_1}\delta(\la_1^{CI}-\la_{n_1}^{BI})\prod_{j_1\neq 1}^{l_1}g(\la_{j_1}^{CI},\la_1^{CI})\prod_{k_1\neq n_1}^{l_1}g(\la_{n_1}^{BI},\la_{k_1}^{BI})\right\}\nonumber\\
&&\times\quad{\lim _{N\rightarrow\infty}}\prod_{CI\neq 1}^{l_1}\prod_{BI\neq{n_1}}^{l_1}\left[{r(\la^B_I)\over{r(\la^C_I)}}\right]^{{N_3\over{2}}+1}\alpha\delta(\la^C_I)^{-{N_1\over{2}}} \alpha\delta(\la^B_I)^{-{N_1\over{2}}}{\cal S}^{(I)}_{l_1}(a(\la)h(\la,\la_{n_1}^{BI}),d(\la)h(\la_1^{CI}))\label{rrd}\nonumber\\
\eeqa
where we have used relation (\ref{delta}).
One should note that the second term on the r.h.s. only contributes - due to the appearance of a delta function - to scattering processes where at least one magnon goes over unscattered from the incoming to the outgoing state. The irreducible scattering amplitude however refers by definition to that part of the amplitude from which all energy conserving subprocesses have been subtracted. It means that the restriction to the irreducible amplitudes effectively implies that only the first term of the recursion relation (\ref{rrd}) has to be taken into account for the evaluation of ${\cal S}^{(I)}$. The ensuing one-term recursion relation is easily solved with the result
\beqa
&&{\lim _{N\rightarrow\infty}}\prod_{CI}^{l_1}\prod_{BI}^{l_1}\left[{r(\la^B_I)\over{r(\la^C_I)}}\right]^{{N_3\over{2}}+1}\alpha\delta(\la^C_I)^{-{N_1\over{2}}} \alpha\delta(\la^B_I)^{-{N_1\over{2}}}{\cal S}^{(I)}_{l_1}(\{\la^C_I\},\{\la^B_I\})_{irr.}= \nonumber\\
&&\hspace{2.5em}\sum_{n_1=1}^{l_1}\sum_{{n_2=1}\atop{\small n_2\neq n_1}}^{l_1}\ldots\sum_{{n_{l_1}=1}\atop{\small n_{l_1}\neq {n_1,\ldots,n_{l_1-1}}}}\prod_{j> i=1}^{l_1}g(\la_i^{CI},\la_j^{CI})\prod_{k_1\neq n_1}^{l_1}g(\la_{k_1}^{BI},\la_{n_1}^{BI})\ldots\prod_{k_{l_1}\neq{n_1,\ldots,n_{l_1}}}^{l_1}g(\la_{k_{l_1}}^{BI},\la_{n_{l_1}}^{BI})\nonumber\\
&&\hspace{1.5em}\times\quad\prod_{k> i=1}^{l_1}\left[h(\la_{k}^{CI},\la_{n_i}^{BI})\,h(\la_{n_k}^{BI},\la_{i}^{CI})\right]\prod_{k=1}^{l_1}\left[{r(\la^B_I)\over{r(\la^C_I)}}\right]^{{N_3\over{2}}+1}g(\la_k^{CI},\la_{n_k}^{BI})\,\,.\label{srr}\nonumber\\
\eeqa
with the subscript ``$irr$'' indicating the restriction to that part of the scalarproduct that contributes finally to the irreducible amplitude.
The extension of the r.h.s. of equation (\ref{srr}) by factors $1={{g(\la,\mu)}\over{g(\la,\mu)}}$ enables us to represent it as a determinant multiplied by some overall factor:
\beqa
&&{\lim _{N\rightarrow\infty}}\prod_{CI}^{l_1}\prod_{BI}^{l_1}\left[{r(\la^B_I)\over{r(\la^C_I)}}\right]^{{N_3\over{2}}+1}\alpha\delta(\la^C_I)^{-{N_1\over{2}}} \alpha\delta(\la^B_I)^{-{N_1\over{2}}}{\cal S}^{(I)}_{l_1}(\{\la^C_I\},\{\la^B_I\})_{irr}= \nonumber\\
&&\hspace{1.5em}\prod_{j>i}^{l_1}g(\la_i^{CI},\la_j^{CI})\prod_{j>i}^{l_1}g(\la_j^{BI},\la_i^{BI})\prod_{i,j}^{l_1}h(\la_i^{CI},\la_j^{BI})\prod_i^{l_1}\left[{r(\la_i^{BI})\over{r(\la_i^{CI})}}\right]^{{N_3\over{2}}+1}det_{l_1}{\cal M}^{(I)}\label{m1}
\eeqa
with ${\cal M}_{ij}^{(I)}={{g(\la_i^{CI},\la_j^{BI})}\over{h(\la_i^{CI},\la_j^{BI})}}$.\\
A similar relation holds for the part denoted with $III$ (here only the second term in the recursion relation of the scalarproduct contributes):
\beqa
{\lim _{N\rightarrow\infty}}\prod_{CIII}^{l_3}&&\prod_{BIII}^{l_3}\left[{r(\la^C_{III})\over{r(\la^B_{III})}}\right]^{{N_1\over{2}}+1}\alpha\delta(\la^C_I)^{-{N_3\over{2}}} \alpha\delta(\la^B_I)^{-{N_3\over{2}}}{cal S}^{(III)}_{l_3}(\{\la^C_{III}\},\{\la^B_{III}\})_{irr}= \nonumber\\
&&\hspace{-4.5em}\prod_{j>i}^{l_3}g(\la_j^{CIII},\la_i^{CIII})\,g(\la_i^{BIII},\la_j^{BIII})\prod_{i,j}^{l_1}h(\la_i^{BIII},\la_j^{CIII})\prod_i^{l_3}\left[{r(\la_i^{CIII})\over{r(\la_i^{BIII})}}\right]^{{{N_1}\over{2}}+1}det_{l_3}{\cal M}^{(III)}\label{m3}
\eeqa
with ${\cal M}^{(III)}_{ij}={{g(\la_j^{BIII},\la_i^{CIII})}\over{h(\la_j^{BIII},\la_i^{CIII})}}\,$.\par
Inserting (\ref{m1}) and (\ref{m3}) into (\ref{mael}) one obtains
\beqa 
&&\hspace{-1.5em}\langle 0|\prod^{l}_{j=1}\cer(\lambda^C_j)\,{\cal O}_{n,n+1}\,\prod^{l}_{k=1}\br(\lambda^B_k)|0\rangle_{irr}= \nonumber \\
&&\prod_{i}^l\left[{r(\la_{i}^{C})\over{r(\la_{i}^{B})}}\right]^{{N_1-N_3}\over{2}}\prod_{j>i}^l g(\la^C_j,\la^C_i)g(\la^B_i,\la^B_j)\sum_{I,II,III}(-1)^{\left[P_C\right]+\left[P_B\right]}\langle 0|\prod_{II}\cer_2(\lambda^C_{II})\,{\cal O}_{n,n+1}\,\prod_{II}\br_2(\lambda^B_{II})|0\rangle\nonumber\\
&&\nonumber\\
&&\times \quad \prod_{I\leq J<K\leq III}h(\la^C_{K},\la^C_{J})\, h(\la^B_{J},\la^B_{K})\,\prod_{i,j}^{l_1}h(\la_i^{CI},\la_j^{BI})\prod_{i,j}^{l_3}h(\la_i^{BIII},\la_j^{CIII})\nonumber\\
&&\times\quad det\left(R^{-1}(\la^C_I){\cal M}^{(I)}(\la^C_I,\la^B_I)R(\la^B_I)\right)det\left(R(\la^C_{III}){\cal M}^{(III)}(\la^C_{III},\la^B_{III})R^{-1}(\la^B_{III})\right)\label{mop} \nonumber\\
\eeqa
with $R(\la)_{ij}=r(\la_i)\delta_{ij}$. While deriving this result we used $f(\la,\mu)=g(\la,\mu) h(\la,\mu)$ and the antisymmetry of the $g$'s.\\
$[P_B]$ stands for the parity of the permutation
\beqan
P_B:\quad\left\{\la^B_{II}\right\}\cup\left\{\la^B_I\right\}\cup\left\{\la^B_{III}\right\}\rightarrow\left\{\la^B\right\}
\eeqan
while $[P_C]$ stands for the parity of the permutation
\beqa
P_C:\quad\left\{\la^C_{II}\right\}\cup\left\{\la^C_I\right\}\cup\left\{\la^C_{III}\right\}\rightarrow\left\{\la^C\right\}
\eeqa
with the enumeration in each subset according to the original one.\\
It is possible in principle to write the result in a more compact way, namely as the determinant of the sum of three matrices \cite{aqf}. As it is not useful for our purpose we will not pursue this line of reasoning.
\section{Low-energy limit}
To start with let us make the simplifications which are due to the special form of the perturbation. The matrixelement
\beq
_2\bra\prod_i \tilde{C}_2(\la^C_i)\vec{\si_n}\vec{\si_{n+1}}\prod_i \tilde{B}_2(\la^B_i)\ket_2\label{mtb2}
\eeq
appearing in equation (\ref{mop}) is to be evaluated with respect to the two--site highest weight state $\ket_2$. There can therefore at most two operators $B$ and $C$ show up in (\ref{mtb2}) (applying two operators $B_2$ to $\ket_2$ one reaches the state of lowest weight of the two-site vector space). Since we are anyhow restricting our attention to scattering events in non-forward direction we may evaluate instead of (\ref{mtb2}) the matrixelement
\beq
_2\bra\prod_i \tilde{C}_2(\la^C_i)\left(\vec{\si_n}\vec{\si_{n+1}}-\id_n\cdot\id_{n+1}\right)\prod_i \tilde{B}_2(\la^B_i)\ket_2\label{mtb2a}
\eeq
with $\id_n$ the identity in $V_n$ (the addition of $\id_n\cdot\id_{n+1}$ gives only a contribution to the amplitude in forward direction).
But (\ref{mtb2a}) vanishes on the state of highest weight (no operators $B_2$ and $C_2$) and on the state of lowest weight (two operators $B_2$ and $C_2$). We are left with the matrixelement with one operator $B_2$ and $C_2$, which is straightforwardly calculated
\beq
_2\bra\tilde{C}_2(\la^C_{II})\left(\vec{\si_n}\vec{\si_{n+1}}-\id_n\cdot\id_{n+1}\right)\tilde{B}_2(\la^B_{II})\ket_2=2\,{1\over{\alpha\delta(\la^B_{II})}} {1\over{\alpha\delta(\la^C_{II})}}\,\,. 
\eeq
Taking the normalization and the last result into account we obtain for the transition amplitude the representation
\beqa
&&\sum_n \langle 0|\prod^{l}_{j=1}Z^{\dagger}(\lambda^C_j)\,{\cal O}_{n,n+1}\,\prod^{l}_{k=1}Z(\lambda^B_k)|0\rangle=\nonumber\\
&&\sum_n\,z_n\,2\prod_{i}^l\left[{r(\la_{i}^{C})\over{r(\la_{i}^{B})}}\right]^{{N_1-N_3}\over{2}}\prod_{j>i}^l {{g(\la^C_j,\la^C_i)}\over{f(\la^C_j,\la^C_i)}}{{g(\la^B_i,\la^B_j)}\over{f(\la^B_i,\la^B_j)}}\sump(-1)^{\left[P_C\right]+\left[P_B\right]}{1\over{\alpha\delta(\la^B_{II})}} {1\over{\alpha\delta(\la^C_{II})}}\nonumber\\
&&\prod_{I\leq J<K\leq III}h(\la^C_{K},\la^C_{J})\, h(\la^B_{J},\la^B_{K})\,\prod_{i,j}^{l_1}h(\la_i^{CI},\la_j^{BI})\prod_{i,j}^{l_3}h(\la_i^{BIII},\la_j^{CIII})\nonumber\\
&&\times\quad det\left(R^{-1}(\la^C_I){\cal M}^{(I)}(\la^C_I,\la^B_I)R(\la^B_I)\right)det\left(R(\la^C_{III}){\cal M}^{(III)}(\la^C_{III},\la^B_{III})R^{-1}(\la^B_{III})\right)\label{er}
\eeqa
The slash on the sum over the partitions is supposed to indicate that only partitions with exactly one representative present in the subset labeled by $II$ are to be taken.\par
We are now prepared to examine the behaviour of irreducible scattering amplitudes at low momenta ($\la_i \sim p_i$ for small momentum) with two or more than two magnons involved (the one  particle amplitude will be quoted below for the sake of completeness). An obvious method to get a handle on formula (\ref{er}) consists in a systematic expansion in powers of momenta, as far as they appear in functions $h$ and keeping at the same time the functions $g$ unexpanded. The leading term is obtained by putting $h$ consistently to one at all places where it appears in (\ref{er}). This yields fot $\la_{\alpha}\,\in\,\{\la^C\}$
\beqa
&&\sum_n \langle 0|\prod^{l}_{j=1}Z^{\dagger}(\lambda^C_j)\,{\cal O}_{n,n+1}\,\prod^{l}_{k=1}Z(\lambda^B_k)|0\rangle\approx\nonumber\\
&&\sum_n 32\,z_n\prod_{j>i}^l {{g(\la^C_j,\la^C_i)}\over{f(\la^C_j,\la^C_i)}}{{g(\la^B_i,\la^B_j)}\over{f(\la^B_i,\la^B_j)}}\sump(-1)^{\left[P_C\right]+\left[P_B\right]}det\,g(\la^C_I,\la^B_I)\,det\,g(\la^B_{III},\la^C_{III})\,\,.\label{1o}\nonumber\\
\eeqa
The prefactors $\prod_{j>i}^l{g(\la_i,\mu_j)\over{f(\la_i,\mu_j)}}$ may also be put equal to one in leading order by noting that ${g(\la,\mu)\over{f(\la,\mu)}}\approx 1+O\left(\la-\mu\right)$.
The sum over the partitions I and III in (\ref{1o}) renders a vanishing result as one infers from the Laplace formula for the determinant of a sum of matrices \cite{bog}:
\beq
det(A+B)=\sum_{P_L,P_C}(-1)^{[P_L]+[P_C]}det\,A_{P_L,P_C}det\,B_{P_L,P_C}
\eeq
where $P_L$ is the partition of rows in subsets of rows of $A$ and $B$, while $P_C$ is analogous the partition of columns, and the fact that $g$ is odd
\beqan
g(\la^C_i,\la^B_j)+g(\la^B_j,\la^C_i)=0\,\, .
\eeqan
For the next order of the expansion in powers of momenta we obtain in a straightforward manner the following result:
\beqa
&&\langle 0|\prod^{l}_{j=1}Z^{\dagger}(\lambda^C_j)\,z_i\vec{\sigma_i}\,\vec{\sigma_{i+1}}\,\prod^{l}_{k=1}Z(\lambda^B_k)|0\rangle\approx 2 i\,\prod_{j>i}^l h^{-1}(\la^C_j,\la^C_i)h^{-1}(\la^B_i,\la^B_j)\prod_{i}\left[{r(\la_{i}^{C})\over{r(\la_{i}^{B})}}\right]^{{N_1-N_3}\over{2}}\nonumber \\
&&\times\quad 16\, z_i\sum_{C}\la_{\alpha}\sump(-1)^{[P_B]+[P_C]}(l_3-l_1+6\epsilon_{\alpha}^{I,III})\,det \,g(\la^{C}_I,\la^B_I)\, det\, g(\la^{B}_{III},\la^C_{III}) \label{r1}
\eeqa
with
\beqan
\epsilon_{\alpha}^{I,III}=\left\{\begin{array}{*{1}{c}} +1;\,\la_{\alpha}\in I\\-1;\,\la_{\alpha}\in III\\0;\,\la_{\alpha}\in II\end{array}   \right. 
\eeqan
If $\la_{\alpha}\in\{\la^B\}$ we get the same result up to an overall minus sign.\\
The result can be simplified further, using the following chain of identities:
\beqan
&&\sum_{I,III}(-1)^{[P_B]+[P_C]}(l_3-l_1)\,det \,g(\la^{C}_I,\la^B_I)\, det\, g(\la^{B}_{III},\la^C_{III})
\nonumber\\
&&={\partial\over{\partial x}}\sum_{I,III}(-1)^{[P_B]+[P_C]}\,det \,g(x\la^{C}_I,x\la^B_I)\, det\, g(x^{-1}\la^{B}_{III},x^{-1}\la^C_{III})|_{x=1}\nonumber\\
&&={\partial\over{\partial x}}\,det\left(g(x\la^{C},x\la^B)+g(x^{-1}\la^{B},x^{-1}\la^C)\right)|_{x=1}\nonumber\\
&&={\partial\over{\partial x}}\left.\left({1\over{x}}-x\right)^{l_1+l_3}\right|_{x=1}det\,g(\la^{C},\la^B)\nonumber\\
&&=0\hspace{2em};{\mbox{for}}\,\,\l_1+l_3>1
\eeqan
where we used again the Laplace formula and the antisymmetry of the $g$'s.\\
The remaining term of the first-order Taylor expansion is 
\beqan
&&\langle 0|\prod^{l}_{j=1}Z^{\dagger}(\lambda^C_j)\,z_i\vec{\sigma_i}\,\vec{\sigma_{i+1}}\,\prod^{l}_{k=1}Z(\lambda^B_k)|0\rangle= \prod_{j>i}^l h^{-1}(\la^C_j,\la^C_i)h^{-1}(\la^B_i,\la^B_j)\prod_{i}\left[{r(\la_{i}^{C})\over{r(\la_{i}^{B})}}\right]^{{N_1-N_3}\over{2}}\nonumber \\
&&\times\quad 16\cdot 6 i\,z_i\,\sum_{C}\la_{\alpha}\,\sump(-1)^{[P_B]+[P_C]}(\epsilon_{\alpha}^{I,III})\,det \,g(\la^{C}_I,\la^B_I)\, det\, g(\la^{B}_{III},\la^C_{III})\label{r2}. 
\eeqan
The sum over the first and the third partition can be combined to a determinant of the sum of two matrices 
\beqan
&&\sum_{I,,III}(-1)^{[P_B]+[P_C]}(\epsilon_{\alpha}^{I,III})\,det \,g(\la^{C}_I,\la^B_I)\, det\, g(\la^{B}_{III},\la^C_{III})\nonumber\\
&&=det\left(g(\la^{C},\la^B)+\hat{g}_{\alpha}(\la^{B},\la^C)\right)
\eeqan
The matrix $\hat{g}_{\alpha}(\la^{B},\la^C)$ differs from $g(\la^B,\la^C)$ in that the $\alpha$-th row is multiplied by $(-1)$. The sum of the two matrices is thus a matrix with only one row of non-vanishing entries. The determinant is again zero except for $l_1 + l_3 =1$. We keep as net result that the first--order term of the Taylor expansion gives a non--vanishing contribution only for the one-- and two--particle amplitude.\\
We turn now to the second-order contribution. The computation is tedious, but it proceeds otherwise along the same lines as the first-order calculation. We thus only quote the result (omitting the prefactors):
\beqa
&&\langle 0|\prod^{l}_{j=1}Z^{\dagger}(\lambda^C_j)\,z_i\vec{\sigma_i}\,\vec{\sigma_{i+1}}\,\prod^{l}_{k=1}Z(\lambda^B_k)|0\rangle \sim\nonumber\\
&&16\,\half\sum_{\alpha\beta}\la_{\alpha}\la_{\beta}\sump(-1)^{[P_B]+[P_C]}\left[i\epsilon_{\alpha}^{B,C}(l_1-l_3-6\epsilon_{\alpha}^{I,III})i\epsilon_{\beta}^{B,C}(l_1-l_3-6\epsilon_{\beta}^{I,III})\right.\nonumber\\
&&\left.+\delta_{\alpha\beta}(34\,\tilde{\epsilon}^{I,III}+l_1+l_3)\right]\,det \,g(\la^{C}_I,\la^B_I)\, det\, g(\la^{B}_{III},\la^C_{III})\label{r20}
\eeqa
with
\beqan
\epsilon_{\alpha}^{B,C}=\left\{\begin{array}{*{1}{c}} +1;\,\la_{\alpha}\in B\\-1;\,\la_{\alpha}\in C\end{array}   \right. 
\eeqan
and
\beqan
\tilde{\epsilon}^{I,III}=\left\{\begin{array}{*{1}{c}} +1;\,\la\,\, {\rm{in}} \,\,I, III\\0;\,\rm{otherwise}\end{array}   \right. 
\eeqan

When $l_1+l_3>2$ this term vanishes as can be shown by generalizing the considerations used with the first-order calculation:
\begin{itemize}
\item terms proportional to $(l_1 + l_3 + const)$ vanish by the same argument as used in the first-order calculation (even for $l_1+l_3=2$)
\item terms proportional to $(l_1-l_3)^2$ vanish when regarded as the second derivative with respect to $x$ at $x=1$ (this term renders for $l_1+l_3=2$ the only nonvanishing contribution)
\item terms proportional to $\epsilon_{\alpha}^{I,III}(l_1-l_3)$ give a matrix with a prefactor $\left({1\over{x}}-x\right)^{l_1+l_3-1}\left({1\over{x}}+x\right)$ of which the derivative with respect to $x$ at $x=1$ vanishes
\item terms proportional to $\epsilon_{\alpha}^{I,III}\epsilon_{\beta}^{I,III}$ give a matrix with at most 2 columns or rows or one column and one row not zero after applying the Laplace formula 
\end{itemize}\par
We finish this section by quoting the leading terms of the transition amplitudes at small momenta with the explicit expressions for $l<3$ in lowest order:
\begin{itemize}
\item $l=1$\\
\beqa
\sum_n z_n\bra Z^{\dagger}(\la)\,\vec{\sigma_n}\,\vec{\sigma_{n+1}}\,Z(\mu)\ket&=&-2 \sum_n\,z_n \left[{r(\la)\over{r(\mu)}}\right]^{{N_1-N_3}\over{2}}{1\over{\la^2 +\quar}}{1\over{\mu^2+\quar}}\label{l1}\nonumber\\
\eeqa
\item $l=2$, cf. equation (\ref{r1})\\
\beqa
\sum_n z_n\bra Z^{\dagger}(\la_1^C)Z^{\dagger}(\la_{2}^C)\,\vec{\sigma_n}\,\vec{\sigma_{n+1}}\,Z(\la^B_{1})Z(\la^B_{2})\ket &=& \nonumber\\
&&\hspace{-7cm}-64 \sum_n z_n \left(\sum_C \la^C-\sum_B\la^B\right)^2\,det_2\left({1\over{\la^C-\la^B}}\right)\label{l2}\nonumber\\
\eeqa
\item $l=3$, cf. equation (\ref{r20})\\
\beqa
\sum_n z_n\bra \prod_{i=1}^3Z^{\dagger}(\la_i^C)\,\vec{\sigma_n}\,\vec{\sigma_{n+1}}\,\prod_{i=1}^3Z(\la^B_{i})\ket &=&\nonumber\\
&&\hspace{-6.5cm}-128\sum_n z_n\left(\sum_C \la^C-\sum_B\la^B\right)^3\,det_3\left({1\over{\la^C-\la^B}}\right)\label{l3}\nonumber\\
\eeqa
\end{itemize}
with $det_l\left({1\over{\la^C-\la^B}}\right)$ denoting the Cauchy determinant of a $l\times l$ matrix
\beqan
det_l\left({1\over{\la_i-\mu_j}}\right)=(-1)^{{l(l-1)\over{2}}}{{\prod_{i<j}^l(\la_i-\la_j)\,\prod_{i<j}^l(\mu_i-\mu_j)}\over{\prod_{i,j}^l(\la_i-\mu_j)}}\,\, .
\eeqan
\par
A couple of remarks may be in order:\\
\hspace*{1cm}1. The one-particle amplitude quoted above is in fact the full Born term (not the leading piece at small momentum).\\
\hspace*{1cm}2. We have left out in (\ref{l2}) and (\ref{l3}) factors $
\prod_i\left[{r(\la_i)\over{r(\mu_i)}}\right]^{{N_1-N_3}\over{2}}$. The latter give rise to a Fourier transform of the distribution of coupling constants. We have effectively evaluated the distribution at zero momentum.\\
\hspace*{1cm}3. To apply the above expressions to physical processes of magnon scattering one has to restrict the respective expressions to the energy shell, given by $\sum_i {\la^B_i}^2=\sum_i {\la^C_i}^2$.

\section{Conclusion}
The main result of this paper are the formulas (\ref{l2}) and (\ref{l3}) for two-- and three--magnon scattering at small momenta. An obvious generalization to n--particle scattering may be conjectured:
\beq
\bra \prod_{i=1}^l Z^{\dagger}(\la_i^C)\,\vec{\sigma_i}\,\vec{\sigma_{i+1}}\,\prod_{i=1}^l Z(\la^B_{i})\ket \approx -16\,2^l\left(\sum_C \la^C-\sum_B\la^B\right)^l\,det_l\left({1\over{\la^C-\la^B}}\right)\,\,.
\eeq
We are not able to prove this conjecture so far.\par
There are other kinematical regions besides the one of low momenta for which simple and reliable estimates can be made. If all momenta and all differences of momenta become large, the n--particle transition amplitude decreases with $\rho ^{-(n+3)}$ -- $\rho$ denoting a common scale of all momenta -- as can be inferred from an inspection of equation (\ref{mop}). An interesting kinematical region -- also accessible to a rather detailed analytical description -- is given by the setting 
\beqan
|\la_i^B-\la^B_j|\ll 1,\quad |\la_i^C-\la^C_j|\ll 1,\quad |\la_i^B-\la^C_j|\gg 1,\,\,\forall\,\,i,j\,\,,|\la^B|\sim|\la^C|\sim\rho\gg 1.
\eeqan
This situation is realized if a bunch of particles travelling approximatively with the same velocity is collectively scattered backwards at the inhomogeneity.\\
The piece of (\ref{mop}) supplying the $\rho$ dependence in this case is given by 
\beqan
&&_2\langle 0|\cer_2(\lambda^C_{II})\,{\cal O}_{n,n+1}\,\br_2(\lambda^B_{II})|0\rangle_2\,\prod_{i,j}^{l_1}h(\la_i^{CI},\la_j^{BI})\prod_{i,j}^{l_3}h(\la_i^{BIII},\la_j^{CIII})\nonumber\\
&&\quad\times\quad det\left({\cal M}^{(I)}(\la^C_I,\la^B_I)\right)det\left({\cal M}^{(III)}(\la^C_{III},\la^B_{III})\right)
\eeqan
for which one easily calculates the scaling behaviour $\rho^{-(n+3)}$.\\
To arrive at this conclusion it is essential to view the determinants in the above formula as derivatives of Cauchy determinants:
\beqan
det{\cal M}_{ij}\approx det_l{1\over{(\la_i-\la_j)^2}}\sim {{\partial}\over{\partial \la^C_1}}\cdots {{\partial}\over{\partial \la^C_n}}\,det_l{1\over{(\la_i-\la_j)}}
\eeqan
\par
A completely open problem within our approach is the treatment of string states. The determination of break--up amplitudes for string states seems to us a particularly challenging problem.\par
{\bf{Acknowledement}}\\
We thank H.M. Babujian for a collaboration at an early stage of this work.

\begin{appendix}
\section{Appendix}
In this appendix we use the result (\ref{mael}) to determine the shift of energy eigenvalues caused by $H_{inh}=\quar\sum^N_{n=1}z_n[\vec{\si}_n\vec{\si}_{n+1}-\id]$ \cite{dipl2, dipl1}.\\
The lowest excitation is generated by flipping one spin $(l=1)$. The solution of the Bethe Ansatz equation is in this case 
\beq
\lambda={{1}\over{2}}\cot{{p_0}\over{2}}{;}\hspace{5em}p_0={{2\pi k}\over{N}}\; ;\hspace{2em}k=1,...,N \,\, .
\eeq
Taking into account parity degeneracy the first-order correction to the energy $E^{\left(0\right)}(\la)=\half{{1}\over{\lambda^2+\quar}}$ is found to be
\beqan
E^{\left(1\right)}={{{\cal V}(\lambda,\lambda)}\over{\langle0|C(\lambda)B(\lambda)|0\rangle}}\pm 2{{|{\cal V}(\lambda,-\lambda)|}\over{\langle0|C(\lambda)B(\lambda)|0\rangle}}
\label{stoerung}
\eeqan\\
with ${\cal V}(\mu_1,\mu_2)=-\quar\sum_{j=1}^N z_j\langle
0|C(\mu_1)(\vec{\sigma}_j\vec{\sigma}_{j+1}-\id)B(\mu_2)|0
\rangle$ which leads to\\
\begin{eqnarray}
&&E^{\left(1\right)}(\lambda)=E^{\left(0\right)}(\lambda)\left[\left({{1}\over{N}}\sum_{j=1}^N z_j\right)\pm \sqrt{\sum_{j,k=1}^N{{z_j z_k}\over{N^2}}\;\exp(-2 i p_0(\lambda)(j-k))}\;\right]\,\,.\nonumber\\\label{energyl1}
&&\nonumber\\
\end{eqnarray}
This shows that the energy correction depends in first-order both on the mean--value of the couplings $\,\overline z={{1}\over{N}}\sum_{j=1}^N\,z_j\,$ and on the Fouriertransform of the distribution ${{1}\over{N}}\sum_{j=1}^Nz_j\,\exp(\pm 2 i p_0 j)$ ($p_0={{1}\over{i}}\ln{{i\,\la+\ihalf}\over{i\,\la-\ihalf}}$).\\
The second-order corrections can be obtained from the secular equation
\beqan
E^{\left(2\right)}_{n}(\lambda)=\sum_{m\neq n}{\tilde{\cal{{V}}}_{nm}\tilde{\cal{{V}}}_{mn}\over{ E^{\left(0\right)}_n  -  E^{\left(0\right)}_m}}
\label{secular}
\eeqan
where the matrixelements are taken with respect to the corrected wave-function in the zeroth  approximation
\beqan
\tilde{\cal{V}}_{nm}&=&\langle c^{\left(0\right)\star}_1\Phi(\la)\pm c^{\left(0\right)\star}_2\Phi(-\la)|H_1|\Phi(\mu)\rangle\\
&=& c^{\left(0\right)\star}_1{\cal {V}}(\la,\mu)\pm c^{\left(0\right)\star}_2{\cal {V}}(-\la,\mu)
\eeqan
with $\cal {V}(\la,\mu)$ defined as in (\ref{stoerung}) and $c^{\left(0\right)}$
being the following expressions
\beqan
c^{\left(0\right)}_1&=&\sqrt{{{\cal {V}}(\la,-\la)\over{2|{\cal {V}}(\la,-\la)|}}}\\
c^{\left(0\right)}_2&=&\sqrt{{{\cal {V}}(-\la,\la)\over{2|{\cal {V}}(-\la,\la)|}}}\,\, .
\eeqan
Inserting the explicit formulas yields
\beqan
 E^{\left(2\right)}_{n}(\lambda)&=&{1\over 4}\sum_{\mu\neq\pm\la}{1\over{\la^2-\mu^2}}{1\over{N^2}}\sum_{j,k}z_j z_k\left\{\left({{i\mu-\ihalf}\over{i\mu+\ihalf}}\right)^{j-k}\left[\left({{i\la+\ihalf}\over{i\la-\ihalf}}\right)^{j-k}+\left({{i\la-\ihalf}\over{i\la+\ihalf}}\right)^{j-k}\right]\right.
\nonumber\\
&&\left.\pm\left({{i\mu+\ihalf}\over{i\mu-\ihalf}}\right)^{j-k}\left[K_{-}\left({{i\la+\ihalf}\over{i\la-\ihalf}}\right)^{j+k}+K_{+}\left({{i\la-\ihalf}\over{i\la+\ihalf}}\right)^{j+k}\right]\right\}
\label{energy2}\,\, .
\eeqan
$K_{\pm}$ is a quotient of Fouriertransforms 
\beqan
K_{\pm}=\sqrt{{{\sum_{j=1}^N z_j\left({{i\la\pm\ihalf}\over{i\la\mp\ihalf}}\right)^{2j}}\over{{\sum_{j=1}^N z_j\left({{i\la\mp\ihalf}\over{i\la\pm\ihalf}}\right)^{2j}}}}}\,\, .
\eeqan
The sum over $\mu$ can be transformed for $N\rightarrow\infty$ into a principle-value integral
\beqan
 E^{\left(2\right)}_{n}(\lambda)&=&{1\over{8\pi}}\sum_{j,k}{{z_j z_k}\over N}\,vp\int_{-\infty}^{+\infty}{1\over{\la^2-\mu^2}}\left\{\left(\mu+\ihalf\right)^{(j-k-1)}\left(\mu-\ihalf\right)^{(k-j-1)}2\cos{[p(\la)(j-k)]}\right.\nonumber\\
&&\pm\left.\left(\mu-\ihalf\right)^{(j-k-1)}\left(\mu+\ihalf\right)^{(k-j-1)}\left[K_{+}e^{-ip(\la)(j+k)}+K_{-}e^{ip(\la)(j+k)}\right]\right\}d\mu \label{integral}\label{pvi}\,\,.
\eeqan
This principle-value integral can be evaluated by deforming the integration contour into the complex plane, closing it at infinity, which is possible as the integrand vanishes as $r^{-4}$ at infinity. Thus only the pole structure of the integral matters.\\
There are poles at $\la=\pm\mu$ for all values of $j,k$, at $\mu=\ihalf$ for $j\ge k$ in the first term  and for $j\le k$ in the second term and at $\mu=-\ihalf$ with $j,k$ dependence of the first and second term interchanged. It is convenient to split the sum over $j,k$ into three parts:
\beqan
\sum_{j,k}=\sum_{j=k}+\sum_{j\ge k}+\sum_{j\le k}\,\, .
\eeqan
For each sum the contour can be deformed in such a way that the integrand only contains poles of first order, for which the residues are easily calculated.

The result of the integration is
\beqa
E^{\left(2\right)}_{n}(\lambda)&=&-E^{0}(\la)\left\{\half\sum_{j}{{z_j^2}\over{N}}\left[2\pm f(\la,j=k)\right]\right.\nonumber\\
&& -\left.{2\over{\la}}\sum_{j>k}{{z_j z_k}\over{N}}\sin{[p(\la)(j-k)]}\left[2\cos{[p(\la)(j-k)]}\pm f(\la,j>k)\right]\right\}\label{energy2l1}
\eeqa
with $f(\la,j,k)=\left[K_{+}e^{-ip(\la)(j+k)}+K_{-}e^{ip(\la)(j+k)}\right]$.
\par
For the second lowest excitation (two magnons) the computation is more involved, but still elementary, so we only give the result for the first-order correction to the energy of the two-magnon state:
\begin{eqnarray*}
&& \hspace{-1.7em}E^{(1)}(\mu_1,\mu_2)=\nonumber\\
&&\hspace{-1.7em} E^{(0)}(\mu_1,\mu_2)
\Bigg\{\sum_{j=1}^N{{z_j}\over N}\pm 2\Bigg[{\sum_{j,k=1}^N{{z_j\,z_k}\over{N^2}}\Big(
\exp\big[-2ip_0(\mu_1)(j-k)\big]+\exp\big[-2ip_0(\mu_2)(j-k)\big]\Big)}
\nonumber\\
&&\hspace{-1.7em}\pm 2 E^{(0)}(\mu_1)E^{(0)}(\mu_2)\Big[{\sum_{j,k=1}^N{{z_j\,z_k}\over{N^2}}\exp\big[
-2ip_0(\mu_1)(j-k)\big]\:\sum_{j,k=1}^N{{z_j\,z_k}\over{N^2}}\exp\big[-2ip_0(\mu_2)(j-k)
\big]}\Big]^{\half}\Bigg]^{\half}\Bigg\}\,\,.
\nonumber\\
\end{eqnarray*}
Furthermore there exist complex solutions of the Bethe Ansatz equations. They describe bound states \cite{bethe} with momentum
\beqan
e^{ip(x)}=\left({{x+i}\over{x-i}}\right) \label{impulsstring}
\eeqan
and energy
\beqan
E_{String}^{0}(x)={1\over{x^2+1}} \label{energiestring}
\eeqan
where $x$ denotes the center of the bound state.\\
The first-order correction for the two magnon bound state is
\begin{eqnarray}
 E_{String}^{\left(1\right)}(x)=E_{String}^{\left(0\right)}(x)\left[\left({{1}\over{N}}\sum_{j=1}^N z_j\right)\pm E_{CM}^{0}\sqrt{\sum_{j,k=1}^N{{z_j z_k}\over{N^2}}\;\exp(-2 i p(x)(j-k))}\;\right]
\end{eqnarray}
with $E_{CM}^{(0)}$ the energy of the center of the bound state. 
\end{appendix}

\end{document}